\documentclass[english,10pt, a4paper, twocolumn,utf8]{article}
\usepackage[T1]{fontenc}
\usepackage[utf8]{inputenc}
\usepackage{babel}
\usepackage{graphicx}
\usepackage[unicode=true,
 bookmarks=true,bookmarksnumbered=true,bookmarksopen=false,
 breaklinks=false,pdfborder={0 0 1},backref=false,colorlinks=false]
 {hyperref}
\hypersetup{pdftitle={DeepIEP: a Peptide Sequence Model of Isoelectric Point (IEP/pI) using Recurrent Neural Networks (RNNs)},
 pdfauthor={Esben Jannik Bjerrum},
 pdfsubject={Deep Peptide Modelling},
 pdfkeywords={Peptides, Isoelectric Point, IEP, pI,predictive modeling, Deep Learning, RNN, LSTM}}

\makeatletter

\providecommand{\tabularnewline}{\\}

\usepackage[]{babel} 
 \usepackage{cite}

\usepackage{amsmath,amsfonts,amsthm} 

\usepackage[svgnames]{xcolor} 

\usepackage[small, labelfont=bf, up, textfont=it]{caption} 

\usepackage{booktabs} 

\usepackage{lastpage} 

\usepackage{graphicx} 

\usepackage{enumitem} 
\setlist{noitemsep} 

\usepackage{sectsty} 
\allsectionsfont{\usefont{OT1}{phv}{b}{n}} 

\usepackage{mdframed} 

\usepackage[font=small,labelfont=bf]{caption}
\makeatletter
\g@addto@macro\@floatboxreset\centering
\makeatother

\usepackage{geometry} 

\usepackage[T1]{fontenc} 
\usepackage[utf8]{inputenc} 

\usepackage{XCharter} 

\usepackage{fancyhdr} 
\newcommand{\headrulecolor}[1]{\patchcmd{\headrule}{\hrule}{\color{#1}\hrule}{}{}}
\newcommand{\footrulecolor}[1]{\patchcmd{\footrule}{\hrule}{\color{#1}\hrule}{}{}}

\renewcommand{\footrulewidth}{1pt} 

\headrulecolor{DarkGoldenrod}
\footrulecolor{DarkGoldenrod}
\fancypagestyle{firstpage}{ 
\renewcommand{\footrulewidth}{0.0pt} 
}

\newcommand{\authorstyle}[1]{{\large\usefont{OT1}{phv}{b}{n}\color{DarkRed}#1}} 

\newcommand{\institution}[1]{{\footnotesize\usefont{OT1}{phv}{m}{sl}\color{Black}#1}} 

\usepackage{titling} 

\pretitle{
\date{}
\vspace{-70pt} 
\vspace{35pt} 
\fontsize{18}{24}\usefont{OT1}{phv}{b}{n}\selectfont 
\color{DarkRed} 
}

\posttitle{\par\vskip 15pt} 

\preauthor{} 

\postauthor{
\vspace{-40pt} 
}

\usepackage{lettrine} 
\usepackage{fix-cm}	

\newcommand{\initial}[1]{ 
\lettrine[lines=3,findent=4pt,nindent=0pt]{
\color{DarkGoldenrod}
{#1}
}{}%
}

\usepackage{xstring} 

\newcommand{\lettrineabstract}[1]{

\mdframed[backgroundcolor=gray!20,hidealllines=true]
\vspace{5pt} 
\StrLeft{#1}{1}[\firstletter] 
\initial{\firstletter}\textbf{\StrGobbleLeft{#1}{1}} 
\vspace{5pt} 
\endmdframed 

\vspace{10pt} 
}

\author{
\authorstyle{Esben Jannik Bjerrum\textsuperscript{1,*}}
\newline\newline 
\textsuperscript{1}\institution{Wildcard Pharmaceutical Consulting, Zeaborg Science Center, Frødings Allé 41, 2860 Søborg, Denmark.}\\ 
\textsuperscript{*}\institution{Corresponding Author: \href{mailto://esben@wildcardconsulting.dk}{esben@wildcardconsulting.dk}} 
}

\makeatother

\begin{document}

\twocolumn[   \begin{@twocolumnfalse}

\title{DeepIEP: a Peptide Sequence Model of Isoelectric Point (IEP/pI) using
Recurrent Neural Networks (RNNs) }

\maketitle
\thispagestyle{fancy}
\renewcommand{\footrulewidth}{0.0pt}
\lhead{}
\chead{}
\rhead{}
\lettrineabstract{The isoelectric point (IEP or pI) is the pH where the net charge on the molecular ensemble of peptides and proteins is zero. This physical-chemical property is dependent on protonable/deprotonable sidechains and their pKa values. Here an pI prediction model is trained from a database of peptide sequences and pIs using a recurrent neural network (RNN) with long short-term memory (LSTM) cells. The trained model obtains an RMSE and R$^2$ of 0.28 and 0.95 for the external test set. The model is not based on pKa values, but prediction of constructed test sequences show similar rankings as already known pKa values. The prediction depends mostly on the existence of known acidic and basic amino acids with fine-adjusted based on the neighboring sequence and position of the charged amino acids in the peptide chain.}

 \end{@twocolumnfalse} ]

\section*{Introduction}

The isoelectric point (pI/IEP) of peptides and proteins is the pH
value where the net charge on the molecular ensemble is zero. It is
of practical interest in work involving peptide and proteins with
regard to solubility/precipitation\cite{Olsen2013}, separation for
proteomics and elution in ion exchange chromatography\cite{Krijgsveld2006,Millioni2013},
in peptide drug design\cite{Fosgerau2015} and for understanding protein
function\cite{SchuurmansStekhoven2008}. Good models has long been
available\cite{Bjellqvist1993}, as the pI can be obtained by merely
counting the number of each type of amino acid, and knowing their
pKa values, solve the Henderson-Hasselbalch equation to determine
the pH where the overall charge is zero. The prediction is dependent
on the pKa values chosen for the amino acid side chains and these
are not the same in a peptide or protein as in bulk solution.\cite{Grimsley2009}
It has been a popular exercise to fine-tune these pKa-sets to obtain
better pI models, and benchmarks of different pKa sets has been conducted\cite{Audain2016,Kozlowski2016}.
However, these models do not take sequence dependent and position
effects into account. As examples, it is known that the clustering
of e.g. acidic residues will lower their apparent pKa as their close
proximity to each others charges will affect the deprotonation and
hence calculation of pI values.\cite{Cargile2008} Additionally, it
has been found that pKa values should be adjusted for amino acids
at the C- or N- terminal, as the close proximity to the amino- or
carboxylic acid from the ends of the main chain will affect the deprotonation
and apparent pKa value. Although it is unclear if this is due to proximity
to the terminal charge\cite{Cargile2008} or formation of artifacts
such as pyroglutamate and pyroaspartate\cite{Lengqvist2011,Zhang2010}.
To potentially solve these effects, specific pKa values for the terminal
amino acids have been employed\cite{Cargile2008} as well as more
advanced machine learning modes using SVMs\cite{Branca2013} and including
descriptors calculated with cheminformatics toolkits\cite{Perez-Riverol2012}.

Having previously good experience with employing recurrent neural
networks (RNNs) with long short-term memory cells (LSTM) to make regression
models of serialized small molecules\cite{Bjerrum2017}, it was obvious
to try and employ the same model architecture to the natural sequences
of amino acids in peptides. Previous work has been done with using
artificial neural networks or recurrent neural networks to model protein
and peptide properties and functions \cite{Liu2017,Mittermayr2008,Wang2016},
but it is to my knowledge the first time this has been tried for prediction
of pI. The model is trained directly from a database of sequences
and their corresponding pI value, without any knowledge about pKa
values of amino acids or the Henderson-Hasselbalch equation.

\section*{Methods}

\subsubsection*{Datasets}

The dataset was combined from two datasets\cite{Gauci2008,Heller2005a}
derived from proteomics and a dataset extracted from Reaxys\cite{Reaxys2016}
as previously described\cite{Bjerrum2017a}, using the plain peptide
sequences without modifications. The peptides in the proteomics studies\cite{Gauci2008,Heller2005a}
were reacted with iodoacetamide and the cysteine characters C was
substituted with Z. The pIs from the proteomics studies were set to
the average of the gel fractions where the peptide was identified.
After combination the datasets were randomly divided into a 90\% train
and 10\% test set.

\subsubsection*{Vectorization}

The amino acid sequences were converted to one-hot encoded vectors,
where each row corresponded to an amino acid. The vectors were expanded
to similar dimensions by padding with ones in a separate row. Before
training the vectors were flipped along the sequence axis, so that
the padding characters were the first and the encoded sequences the
last. The amino acid sequence was thus flipped in the same operation,
and the RNN network read the sequence from the C-terminal to the N-terminal.

\subsubsection*{Neural Network Modeling}

An RNN model was constructed using Keras v. 2.1.1\cite{chollet2015}
and Tensorflow v. 1.4\cite{Abadi2016b}. The first layer consisted
of 200 LSTM cells\cite{Hochreiter1997}, with a recurrent dropout
of 0.3 and used in batch mode. The final internal C state was used
as input to a dense layer of 100 neurons with the rectified linear
unit activation function\cite{Nair2010}. The output from the dense
layer was connected to a single output neuron with a linear activation
function.

The network was trained with mini-batches of 256 sequences for 300
epochs using the mean square error loss function and the Adam optimizer
\cite{Kingma2014} with an initial learning rate of 0.005. The loss
was followed on the test set and the learning rate lowered by a factor
of two when no improvement in the test set loss had been observed
for 15 epochs. Training took approximately a quarter of an hour.

After training in batch mode, a stateful model was constructed by
creating a model with the exact same architecture but the LSTM states
set to stateful and the input vector reduced to a size of one in the
sequence dimension. After creation of the stateful model, the weights
for the networks were copied from the batch model.

All computations and training were done on a Linux workstation (Ubuntu
Mate 16.04) with 32 GB of ram, i5-2405S CPU @ 2.50GHz and an Nvidia
Geforce GTX 1060 graphics card with 6 GB of ram.

\section*{Results}

\begin{figure}
\caption{\label{fig:one-hot}Example of one-hot vectorization of the first
sequence of the training set. The sequence has been inverted and the
padding comes before the sequence. The N-terminal Alanine (A) is seen
in the upper right corner, and the padding character is seen as a
line in the lower left corner.}

\includegraphics[width=1\columnwidth]{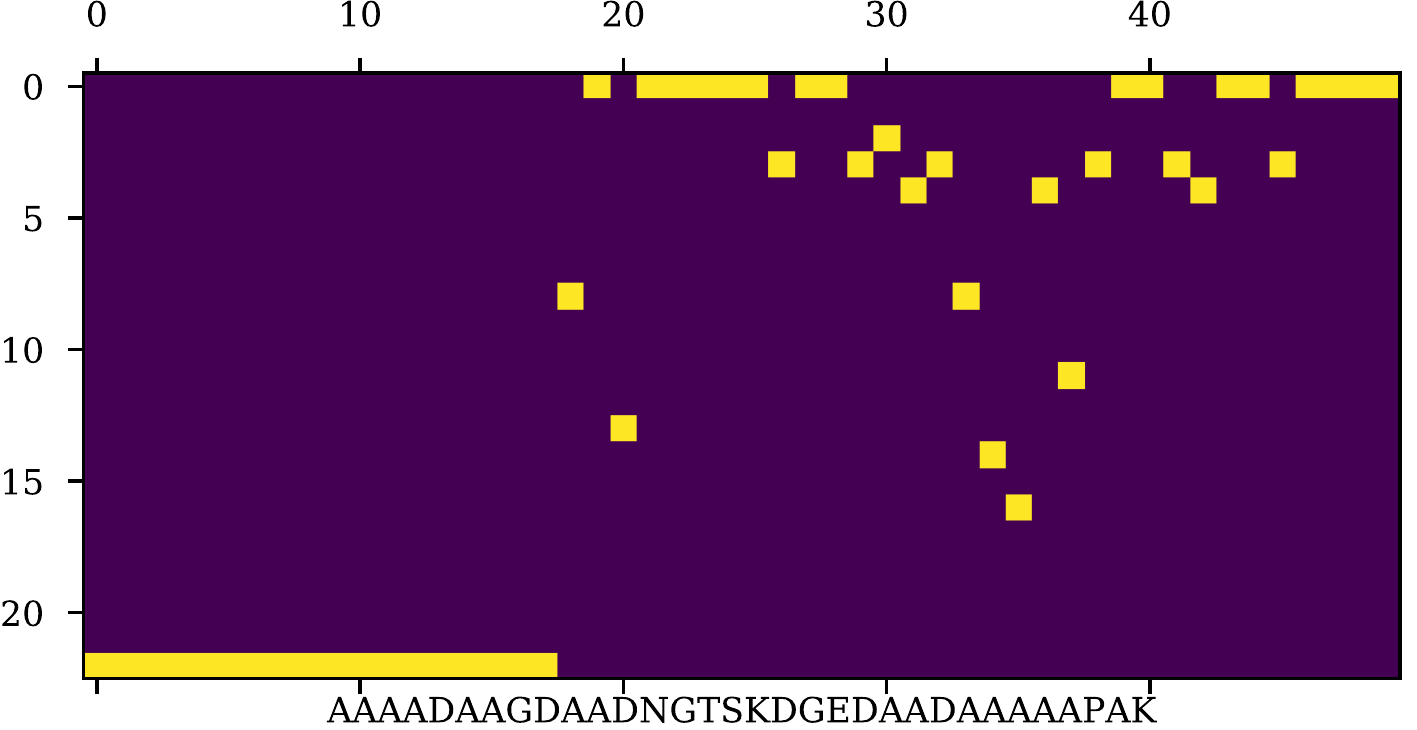}
\end{figure}

Loading and vectorization of the pI dataset resulted in vectors with
the shapes and (7364, 50, 23) and (734, 50, 23) for the train and
test set. The shape is Samples, Sequence, Features (amino acids +
padding). An example of a one-hot encoded peptide is shown in Figure
\ref{fig:one-hot}.

\begin{figure}
\caption{\label{fig:Training-history}Training history of the neural network
model. The loss and test set loss both drops and plateaus off as the
training progresses. Lowering of the learning rate is visible towards
the end of training as periodic lowering of the volatility and/or
loss.}

\includegraphics[width=1\columnwidth]{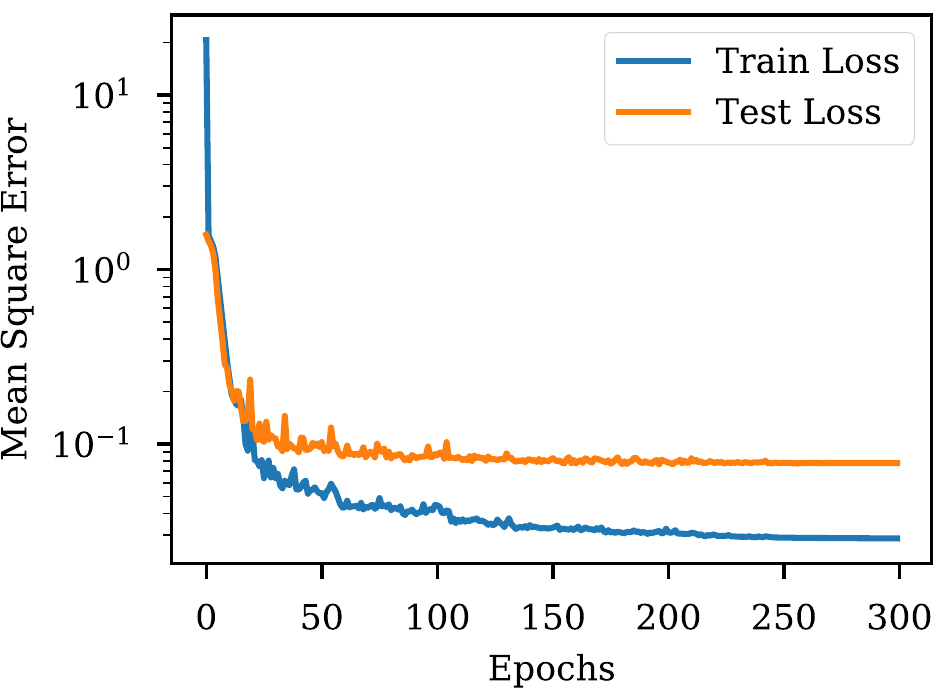}
\end{figure}

The training loss and validation loss improved and plateaued out over
the course of training as shown in Figure \ref{fig:Training-history}.
 There is no sign of over training as there is no rise in the validation
loss function towards the end of training. The final losses were 0.029
and 0.078, for the training and test (validation) loss, respectively.
The spatial difference between the final loss on the test and training
sets appear enlarged due to the logarithmic Y-axis, but makes it easier
to visually see that the training have fully converged.

\begin{figure*}
\caption{\label{fig:Scatter-plot}Scatter plot of literature pI vs. predicted
pI for the training and test set (blue and orange, respectively).
The ideal x=y line is shown in light gray.}

\includegraphics[width=1\textwidth]{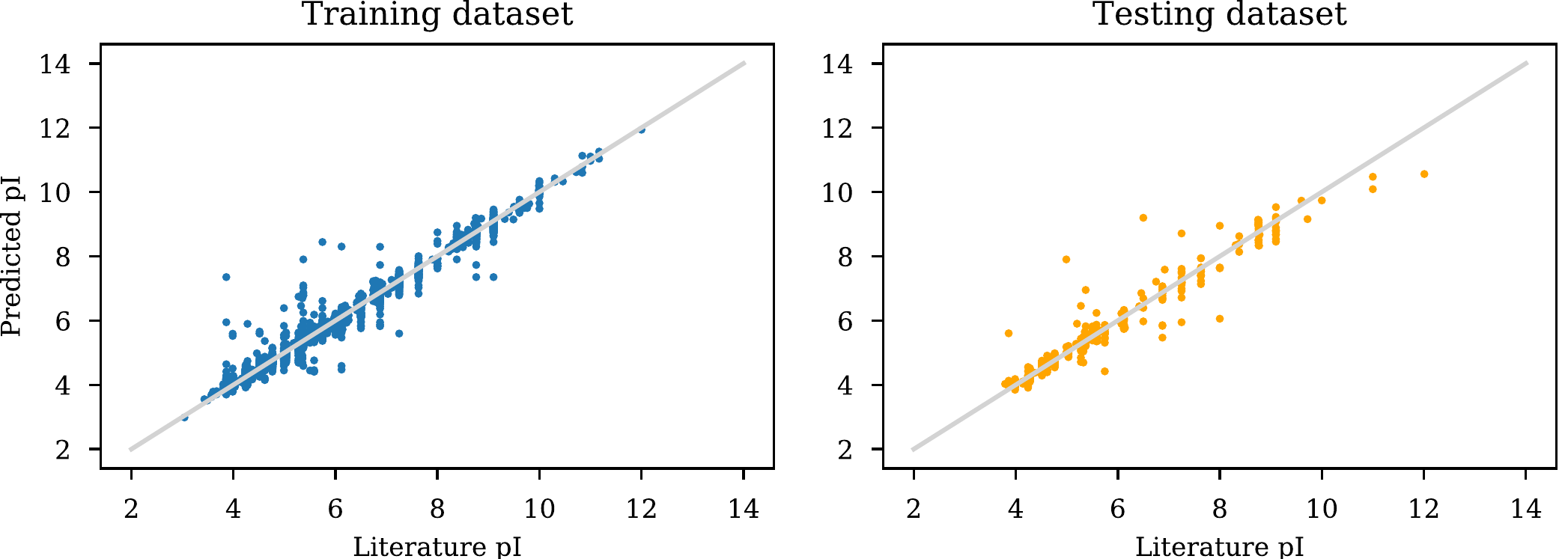}
\end{figure*}

The pIs of the training and test sequences were predicted with the
final model and used in scatter plots and for calculating the root
mean square error (RMSE) and correlation coefficient (R2). (Figure
\ref{fig:Scatter-plot} and Table \ref{tab:Root-mean-square}).
\begin{table}
\caption{\label{tab:Root-mean-square}Root mean square error (RMSE) and correlation
coefficients (R\textsuperscript{2}) of the prediction on the datasets.}

\begin{tabular}{ccc}
\hline 
Dataset & RMSE & R\textsuperscript{2}\tabularnewline
\hline 
Train & 0.17 & 0.98\tabularnewline
Test & 0.28 & 0.95\tabularnewline
\hline 
\end{tabular}
\end{table}

\subsubsection*{Challenging with constructed sequences}

The model was challenged with designed peptide sequences in the hope
to elucidate the effects of amino acid type, position and neighbor
sequence. The normal acidic and basic amino acids was put in the middle
of an alanine peptide sequence and the pI predicted. The predicted
pI is compared to the side chains pKa value in Table \ref{tab:aa_type}.
\begin{table}
\caption{\label{tab:aa_type}Influence of amino acid type on pI prediction.}

\begin{tabular}{ccc}
\hline 
Sequence & pKa\cite{Hass2015} & predicted pI\tabularnewline
\hline 
AAAAA\textbf{D}AAAAA & 4 & 4.23\tabularnewline
AAAAA\textbf{E}AAAAA & 4.4 & 4.35\tabularnewline
AAAAA\textbf{H}AAAAA & 6.8 & 7.05\tabularnewline
AAAAA\textbf{C}AAAAA & 8.3 & 6.08\tabularnewline
AAAAA\textbf{K}AAAAA & 10.4 & 8.35\tabularnewline
AAAAA\textbf{R}AAAAA & 13.5 & 8.82\tabularnewline
\hline 
\end{tabular}
\end{table}
 Standard one-letter amino acid codes are used. The predicted pI should
be a balance between the N- and C- terminal groups and the acidic
and basic side chains. The changes in pI observed, largely follows
the ranking observed for the pKa values, with cysteine (C) as an outlier. 

Table \ref{tab:term_pos} shows the effect of placing the amino acids
in the N and C-terminal respectively, using glutamate (E) and lysine
(K) as examples of an acidic or basic amino acid respectively.
\begin{table}
\caption{\label{tab:term_pos}Influence of positioning at the sequence terminals}

\begin{tabular}{cc}
\hline 
Sequence & predicted pI\tabularnewline
\hline 
AAAAA\textbf{E}AAAAA & 4.35\tabularnewline
\textbf{E}AAAAAAAAAA & 4.11\tabularnewline
AAAAAAAAAA\textbf{E} & 4.52\tabularnewline
\hline 
AAAAA\textbf{K}AAAAA & 8.35\tabularnewline
\textbf{K}AAAAAAAAAA & 8.65\tabularnewline
AAAAAAAAAA\textbf{K} & 8.46\tabularnewline
\hline 
\end{tabular}
\end{table}
 The expectation is that the proximity to the N terminal amino group
or C terminal carboxylic acid moiety, would affect the observed pKa
and thus charge of the side chain either up and down with subsequent
adjustments of the predicted pI. For the glutamate the pI is lowered
when placed in the N-terminal end and higher when placed in the C-terminal
end (to the right). The proximity between the carboxylic acid moiety
of the and the N-terminal ammonium groups positive charge, should
make deprotonation more likely at lower pH, and thus lower the pI
as also observed. The effect is opposite in the other end as also
expected. For the ammonium group of lysine, the effect is reversed.

Some examples of the effect of clustering charged amino acids on the
predicted pI is shown in Table \ref{tab:Effect-of-clustering}.
\begin{table}
\caption{\label{tab:Effect-of-clustering}Influence of clustering charged amino
acids}

\begin{tabular}{cc}
\hline 
Sequence & predicted pI\tabularnewline
\hline 
AAAA\textbf{DDD}AAAA & 3.66\tabularnewline
AA\textbf{D}AA\textbf{D}AA\textbf{D}AA & 3.68\tabularnewline
\hline 
AAAA\textbf{KKK}AAAA & 8.57\tabularnewline
AA\textbf{K}AA\textbf{K}AA\textbf{K}AA & 9.22\tabularnewline
\hline 
AAAA\textbf{RRR}AAAA & 10.78\tabularnewline
AA\textbf{R}AA\textbf{R}AA\textbf{R}AA & 10.92\tabularnewline
\hline 
A\textbf{E}A\textbf{E}AAAA\textbf{K}AA & 4.58\tabularnewline
A\textbf{E}AAA\textbf{EK}AAAA & 4.56\tabularnewline
\hline 
\end{tabular}
\end{table}
 The argumentation is as above, that the proximity of the same charged
amino acids will diminish their acid/base strength. Clustering the
aspartate (D) side chains make the predicted pH slightly higher, where
the lowering of the predicted pI for the basic amino acids (K and
R) is more pronounced. Placing opposite charged side chains besides
each other, enhances the effect of the one closest to the predicted
pI as also observed in the two bottom rows of Table \ref{tab:Effect-of-clustering}.

\subsubsection*{Prediction in stateful mode}

\begin{figure}
\caption{\label{fig:Prediction-profile}Prediction profile obtained with a
stateful model. As the model is presented with different amino acids
encoded as one dimensional one-hot encoded vectors, the predicted
pI changes. The stippled gray line is the Literature pI value of the
full peptide. }

\includegraphics[width=1\columnwidth]{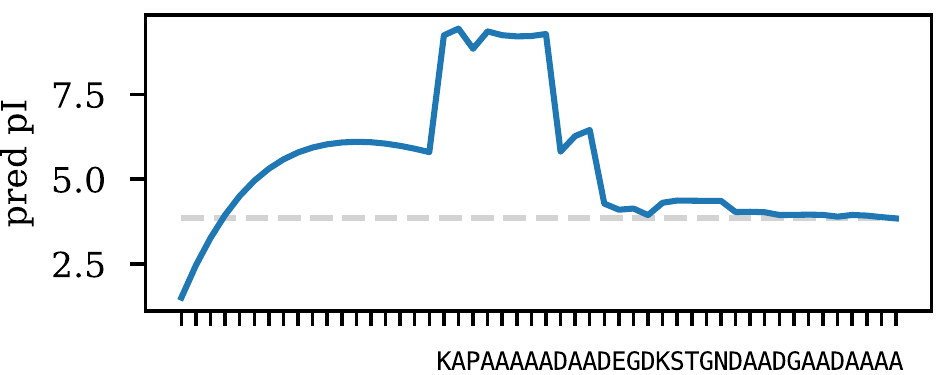}
\end{figure}

The batch trained model was converted into a stateful model, which
do not predict on a sequence, but rather predicts the pI by being
presented one amino acid at a time, with the current prediction being
affected by the previous predictions via the internal states of the
LSTM cells. This allows a prediction profile to be obtained by iterating
over the amino acids in a sequence and following the predicted pI.
An example of such a prediction profile is shown in Figure \ref{fig:Prediction-profile}.
 An initial adjustment is observed in the padding phase, but as soon
as an amino acid is encountered (here lysine, K), the pI jumps up.
The pI is largely kept until an acidic aspartate (D) is encountered.
The next aspartate leads to a further drop in pI, but not as pronounced.
After this adjustment the pI is largely unaffected by the remainder
of the encountered acidic aspartates or amino acids without pronounced
acid/base properties.

\begin{figure}
\caption{\label{fig:Padding_error}Effect of removing the padding before prediction
of the same peptide. The test sequence is the same as in Figure \ref{fig:Prediction-profile}.}

\includegraphics[width=1\columnwidth]{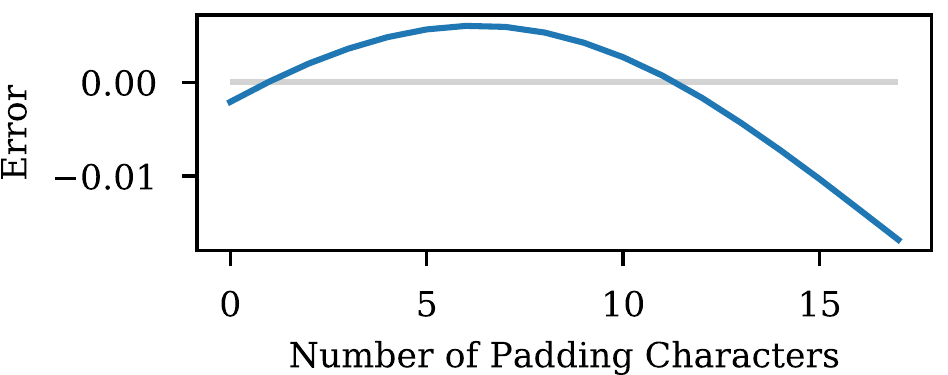}
\end{figure}

The initial changes observed during the padding phase, suggests that
the prediction could be influences by the length of the padding (first
part of Figure \ref{fig:Prediction-profile}). The model was used
for prediction of the same peptide sequence with varying lengths of
padding. The result is shown in Figure \ref{fig:Padding_error}.
 It has a clear effect on the prediction, suggesting that the model
has fine-tuned the prediction based on the size of the peptide, although
for this example, it would have been beneficial to remove the padding
sequence. The error is however in the low range on the second decimal
of the pI prediction.

\section*{Discussion}

Benchmarks of pI calculation pKa sets and pI prediction algorithms
find RMSEs in the range 0.21 to 2.0\cite{Audain2016,Kozlowski2016}.
However, comparison of the datasets published used by Koslowski\cite{Kozlowski2016}
on the website\cite{IPC_homepage}, with the original data\cite{Gauci2008,Heller2005a},
show that he used the predicted pI from the supplementary information
and not the experimentally related average of the gel fractions, which
invalidates the hard work with the benchmark and the tuning of pKa
values. An advanced SVM model obtains RMSE on a test set of 0.21\cite{Audain2016},
but the SVM model had originally\cite{Perez-Riverol2012} been tuned
on part of the dataset that comprised the benchmark, and the performance
is likely overestimated. The originally RMSE found was 0.32\cite{Perez-Riverol2012}.
It thus appears that the approach using RNNs with LSTM cells reach
a performance at least on par with previous methods.

The effect on apparent pKa of glutamate placed at the termini was
studied by Cargile and coworkers, who used a genetic algorithm to
adjust pKa values based on proximity to the terminals and other charged
residues. They found corrections in the range -0.414 for the N-terminus
and 0.014 for the C-terminus. However, this can't be directly compared
to the pI shift of -0.24 and 0.17 observed with the RNN (c.f. Table
\ref{tab:term_pos}), as one is calculated on the pKa level and the
other directly on the predicted pI. The trend is nevertheless the
same and it appears that the neural network has implicitly incorporated
the same necessary correction due to the position of the glutamate
residue.

A problem with the use of data driven approach could be that the experimental
values in databases are not of relevance. An example could be that
iodoacetamide peptide sequences are reported and stored as if they
had free cysteine side chains\cite{Cargile2008,Gauci2008,Heller2005a,Kozlowski2016}.
Another common problem, seem to be that the databases gets polluted
by predicted values from existing models, as it is not always clear
in scientific literature if the pI is predicted or measured. This
could likely pose a problem for a data driven approach as employed
here. However, the model obtained from the data driven approach would
then represent a consensus model of several different models and tuned
pKa sets employed, which could lead to an overall better performance
of the new model due to meta-effects. If only one method had been
employed to derive the pIs for the data driven approach, the new model
would surely regress towards the existing model entirely. Some of
the dataset is also derived from proteomics studies\cite{Gauci2008,Krijgsveld2006,Heller2005a},
where the data has also been filtered with regard to the position
in the isoelectric focusing gel and its deviance from the theoretically
predicted value. This also introduces a bias towards the existing
models, which could likely improve the results, as this filtering
will affect both the training and test set. If the dataset is polluted
by predicted pI values and/or filtered based on predicted values real
world performance of the algorithm would likely be lower than here
observed.

Being a data driven approach, rather than a model based approach,
the method could likely be sensitive to peptide sequence distributions
and unusual local sequences. The database consists of peptides measured
with isoelectric focusing and results from proteomics, which are probably
mostly of eukaryotic and procariotic origin. If peptides following
a different amino acid distribution or being of very low homology
to the ones in the training set (such as from Archea or constructed
peptides) the prediction error could likely be higher. However, as
the model seem mostly influenced by the amino acids with known acid-base
properties (c.f. Figure \ref{fig:Prediction-profile}), the rise in
error could possibly be low. This error could maybe be estimated by
division of the whole peptide dataset into a training and test set
based on clusters found by clustering by sequence homology and redoing
the training and testing.

A better curated or un-biased dataset could likely lead to an increase
in model accuracy and maybe precision. Other improvements could be
the inclusion of modifications if enough data could be collected.
For pIs of peptides with phosphorylated threonines/serines and acetylated/amidated
terminals, this data seems available\cite{Gauci2008}.

\section*{Conclusion}

The feasibility of data driven modeling of peptide isoelectric point
with RNNs has been demonstrated. The model obtained seems on par with
literature sources, illustrating the promise of the approach. However,
the performance of the model is likely overestimated due to shortcomings
of the underlying dataset. The model is mostly influenced by the existence
of amino acids with known acidic and basic properties with minor corrections
due to position in the chain and nearby amino acids. Running the model
in a stateful mode gives a minor error when removing the padding sequence.
A release of the source code and trained weights on the authors GitHub
profile is planned.

\section*{Conflict of interests}

E. J. Bjerrum is the owner of Wildcard Pharmaceutical Consulting.
The company is usually contracted by biotechnology/pharmaceutical
companies to provide third party services. 

\bibliographystyle{elsarticle-num}
\bibliography{Litterature/Litteratur_abbreviated}

\end{document}